\documentclass[a4paper]{article}
\usepackage{indentfirst}
\usepackage{hyperref}
\usepackage{multirow}
\usepackage{latexsym,bm,amsmath,amssymb}
\usepackage{graphicx}
\usepackage{epsfig}
\usepackage{subfigure}
\usepackage{cite}
\usepackage{color}
\usepackage[top=1in, bottom=1in, left=1.25in, right=1.25in]{geometry}
\usepackage{slashed}
\usepackage{fancyhdr}
\usepackage{slashed}
\usepackage{makecell,rotating,multirow,diagbox}
\usepackage{tikz}
\usepackage{float}
\usepackage{appendix}
\usepackage{setspace}
\usepackage{geometry}
\usepackage{graphics}
\usepackage{multirow}
\usepackage{booktabs}
\usetikzlibrary{shapes.geometric, arrows}
\usetikzlibrary{trees}
\usetikzlibrary{decorations.pathmorphing}
\usetikzlibrary{decorations.markings}
\tikzset{
        photon/.style={decorate, decoration={snake}, draw=red},
        nucleon/.style={draw=black, postaction={decorate},
           decoration={markings,mark=at position .55 with{\arrow[draw=black]{>}}}},
        pion/.style={draw=blue, postaction={decorate},
        decoration={markings,mark=at position .55 with{\arrow[draw=blue]{}}}},
        sigma/.style={draw=black, postaction={decorate},
        decoration={markings,mark=at position .55 with{\arrow[draw=black]{}}}},
        link/.style    = { draw=black, double = white, line width = 1.8pt, double distance = 0.8pt,
        postaction={decorate},decoration={markings,mark=at position .55 with{\arrow[draw=black]{>}}}},
    }

\newcommand{\be}{\begin{equation}}
\newcommand{\ee}{\end{equation}}
\newcommand{\ba}{\begin{array}{c}} \newcommand{\ea}{\end{array}}
\newcommand{\bqa}{\begin{eqnarray}}
\newcommand{\eqa}{\end{eqnarray}} 

\def\bea{\arraycolsep .1em \begin{eqnarray}}
\def\eea{\end{eqnarray}}

\def\s0#1#2{\mbox{\small{$ \frac{#1}{#2} $}}}
\def\0#1#2{\frac{#1}{#2}}

\begin{document}
\setcounter{topnumber}{10}
\setcounter{totalnumber}{50}
\title{The $X(6900)$ peak could be a molecular state}
\maketitle
%%%%%%%%%%%%%%%%%%%%%%%%%%%%
%%%%%%%%%%%%%%%%%%%%%%%%%%%%
\begin{center}
{\sc
Ye Lu$^{\dagger\,}$,\,
Chang Chen$^{\dagger\dagger\,}$,\,
Kai-Ge Kang$^{\dagger\dagger\,}$,\,
Guang-you Qin$^{\dagger\,}$,\,
Han-Qing~Zheng$^{\heartsuit\,}$}%\,\footnote{Corresponding author.}
\\
\vspace{0.5cm}
\noindent{\small{$^\dagger$ \it  Institute of Particle Physics and Key Laboratory of Quark and Lepton Physics (MOE),
Central China Normal University, Wuhan, Hubei 430079, China}}\\
\noindent{\small{$^{\dagger\dagger}$ \it  Department of Physics and State Key Laboratory of Nuclear Physics and Technology,
Peking University, Beijing 100871, China}}\\
\noindent{\small{$^\heartsuit$ \it College of Physics, Sichuan University, Chengdu, Sichuan 610065,  China}}\\
\end{center}

\begin{abstract}
The analyses of the  LHCb data on $X(6900)$ found in  di-$J/\psi$ and $J/\psi \psi(3686)$ systems
are performed using a momentum-dependent  {Flatt\'{e}}-like parameterization.
The use of the pole counting rule and spectral density function sum rule give consistent conclusions that $X(6900)$ may not be a molecule of $J/\psi \psi(3686)$. Nevertheless it is still  possible  that $X(6900)$ be a molecule of  higher states, such as
$J/\psi\psi(3770)$, $\chi_{c0}\chi_{c2}$, etc.
\end{abstract}

%

%\textbf{Introduction}
%\label{intro}

The LHCb Collaboration has observed a structure named as  $X(6900)$ in the di-$J/\psi$ invariant mass spectrum~\cite{6900}, with the signal statistical significance  above 5$\sigma$. It is probably composed of four (anti)charm quarks ($c\bar{c} c\bar{c}$) and its width~\cite{6900} are determined to be $80\pm19~({\rm stat.})\pm33~({\rm sys.})$ and $168\pm33~({\rm stat.})\pm69~({\rm sys.})$ MeV in two fitting scenarios of  Breit-Wigner parameterizations with  constant widths. Additionally,  a broad bump and a narrow bump exist in the low and high sides of the di-$J/\psi$ mass~\cite{6900}, respectively, where the former might be a result from a lower broad resonant state (or several lower states)  or interference effect, and the latter is found to be a hint of a state located at $\sim$7200 MeV, named as $X(7200)$.
The $X(6900)$ peak in di-$J/\psi$ channels is also found by the CMS Collaboration~\cite{CMS}.
More recently the $X(6900)$ peak is reported to be observed in the $J/\psi\psi(3686)$ invariant mass spectrum~\cite{ATLAS:2022hhx}.

The experimental  observation has triggered tremendous studies, see for example Refs.~\cite{QFCao} --\cite{Mutuk:2022nkw}. \footnote
{For an incomplete list of earlier studies, see for example  references listed in Ref.~\cite{QFCao}. }
Generally speaking, a molecular state may locate near the threshold of two  color singlet hadrons, like deuteron, $Z_b(10610)$~\cite{BelleZb}, $Z_c(3900)$~\cite{BESIII3900,Belle3900,CLOE3900}, $P_c(4470)$~\cite{Pc2015,Pc2019}, $etc.$
The $X(6900)$ state is close to the threshold of $J/\psi \psi(3770)$, $J/\psi \psi_2(3823)$, ${J/\psi\psi_3(3842)}$, and $\chi_{c0} \chi_{c1}$;
and the $X(7200)$ is close to the threshold of  $J/\psi \psi(4160)$ and $\chi_{c0}$$\chi_{c1}(3872)$.
Inspired by this, it is studied in this paper, as an extension to the work of Ref.~\cite{QFCao}, the properties of $X(6900)$ and $X(7200)$, by assuming for example the $X(6900)$ coupling to $J/\psi J/\psi$, $J/\psi\psi(2s)$, $J/\psi\psi(3770)$, $J/\psi \psi_2(3823)$, ${J/\psi \psi_3(3842)}$ and {$\chi_{c0}\chi_{c2}$} channels, etc.;
and  $X(7200)$ to  $J/\psi J/\psi$, $J/\psi\psi(2s)$ and $J/\psi\psi(4160)$ channels, etc. Notice that in this paper we limit ourselves discussing only the $S$-wave ($l=0$) couplings, since only in this situation we are able to distinguish a molecular state from an `elementary' or a confining state.  Such assignment corresponds to $J^P=0^{++}$, $2^{++}$ spin quantum numbers of $X(6900)$.
For the $S$-wave $J/\psi J/\psi$ coupling, the pole counting rule (PCR)~\cite{pole}, which has been applied to the studies of ``$XYZ$" physics in Refs.~\cite{Zhang:2009bv,Dai:2012pb,X3900,Cao:2019wwt}, and spectral density function sum rule (SDFSR)~\cite{X3900,Baru:2003qq,Weinberg,Weinberg:1965zz,Kalashnikova:2009gt}   are employed to analyze the nature of the two structures in Ref.~\cite{QFCao}. It is found that the di-$J/\psi$ data alone is not enough to judge the intrinsic properties of the two states. It is also pointed out that the $X(6900)$ is unlikely a molecule of $J/\psi \psi(3686)$~\cite{QFCao} -- a conclusion drawn before the new data from Ref.~\cite{ATLAS:2022hhx}.

In this paper we make an upgraded analysis on $X(6900)$ by using the new  $J/\psi\psi(3686)$ data. What we conclude from  this reanalysis is that, even though the $X(6900)$ is very unlikely a molecule of  $J/\psi\psi(3686)$, it is not necessarily an ``elementary state", i.e., a compact $\bar c\bar ccc$ tetraquark state.
Considering that the tetra-quark idea is rather {attractive in the literature (see for example, refs.~\cite{Mutuk:2021hmi}, \cite{Wang:2021kfv}, \cite{Esposito:2021ptx}, \cite{Liu:2021rtn}, \cite{Mutuk:2022nkw} )}, it is  important to make a more careful re-analysis on the nature of $X(6900)$. The present  analysis  points out that, there still exists the possibility that the $X(6900)$ state be a molecular state composed of  particles which thresholds are closer to $X(6900)$ comparing with   $J/\psi\psi(3686)$, such as $J/\psi\psi(3770)$, $J/\psi \psi_2(3823)$, ${J/\psi \psi_3(3842)}$  and $\chi_{c0}\chi_{c2}$, etc.

We start with a two channel re-analysis (i.e., di-$J/\psi$,  $J/\psi\psi(3686)$), using the neural network program which has been recently developed in the study of `XYZ' states in Ref.~\cite{Chen:2022shj}. Then we make a further numerical three channel studies by including another nearby threshold and point out that a strong coupling to the third channel is not excluded by the current data, and hence the $X(6900)$ may still be of molecule nature.

\textbf{Examination of the couple channel situation by using neural network}

For the purpose,  a supervised learning scheme is adopted as in Ref.~\cite{Chen:2022shj}. It means that we should get some `molecule' type and `elementary' type samples beforehand. Then these samples are taken into the machine learning program. The samples named `molecule' are with label `0' and named `elementary' are with label `1'. Our purpose is to let program find out the different characteristics of these input line shapes between `0' and `1'.
We use PCR~\cite{pole} as our criteria. That is, if there exists one pole near one threshold in $s$ plane, the S-wave resonance can be interpreted as the molecule in that channel. On the other side, if there exists one more nearby pole (on different sheets), the $S$-wave resonance is labeled as `elementary'. This method has been used in many works to study the nature of exotic
hadron states, see for example, ~Refs.~\cite{X3900, Cao:2019wwt, Zhang:2009bv, Chen:2021tad}.

To simulate a couple channel resonance amplitude, we chose the Flatt\'e-like parametrization to generate training data. For a  two channel parametrization of invariant mass spectrum, it is written as
\begin{equation}
    \begin{aligned}
    \mathcal{R}_i(s) & =\rho_i(s)\left|\frac{e^{i \phi_i}}{s-M^2+i M\left(g_1 \rho_1(s)+g_2 \rho_2(s)\right)}+b g_{1i}\right|^2 \\
    & +\rho_i(s)bg_{2i},
    \end{aligned}
\end{equation}
where $\rho_i$   ($i = 1,2$) represents two-body phase space factor for final state 1 (FS1) and final state 2 (FS2), respectively. Parameter $M$ is the bare mass of the resonance. Coherent and incoherent background contributions are
also considered as the noise which should be  weak in order not  to influence the judgement on lineshapes. In the region we care about, the background terms, $bg_{1i}$ and $bg_{2i}$ are all first order polynomials of $\sqrt{s}$.
All the parameters are tuned for generating a resonance near FS2 threshold and there is one or two poles near FS2 representing `0' and `1' samples, respectively. In practice, the coupling constants $g_1, g_2$ are important to lineshapes as well as the pole positions in the complex s plane~\cite{Chen:2022shj}.

\begin{figure}[h]
    \centering
    \includegraphics[scale = 0.3]{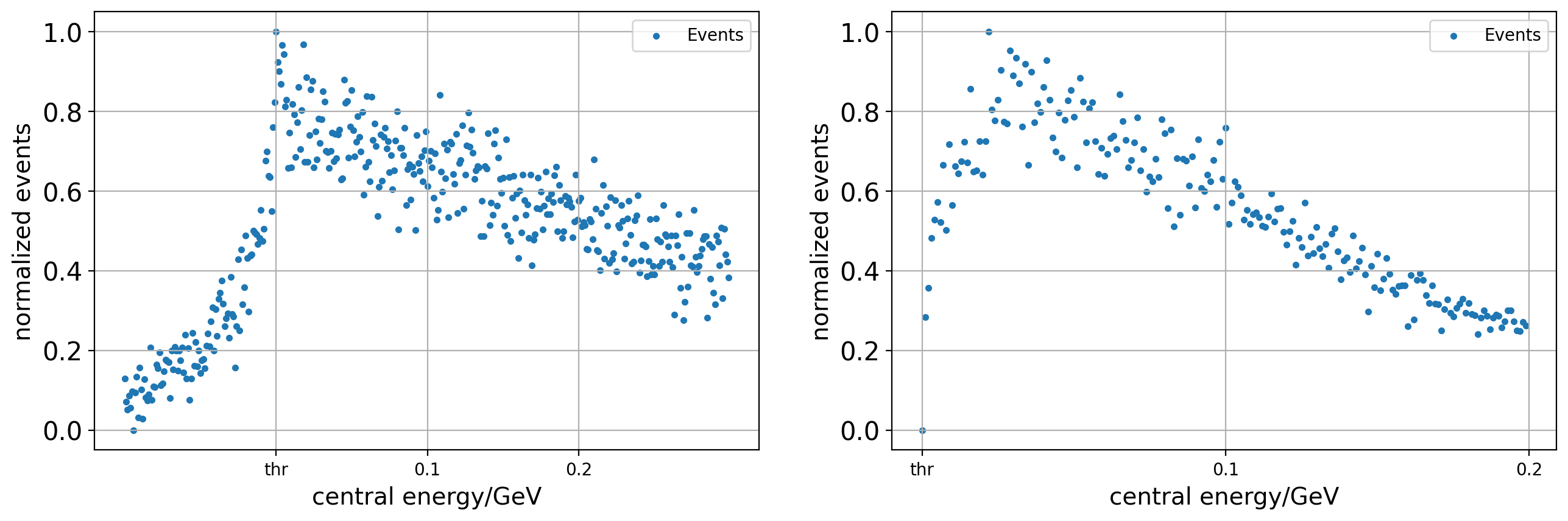}
    \caption{An example of `0' signal from training data for 2 channel case.}\label{ff1}
\end{figure}

\begin{figure}[h]
    \centering
    \includegraphics[scale = 0.3]{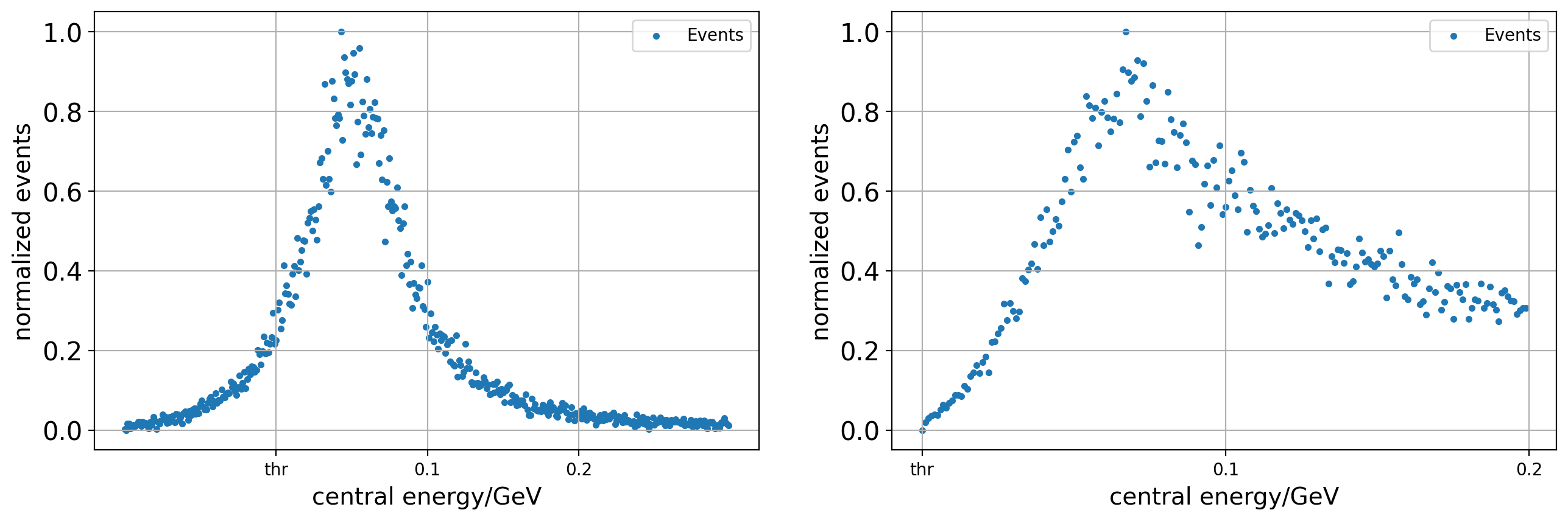}
    \caption{An example of `1' signal from training data for 2 channel case.}\label{ff2}
\end{figure}

\begin{figure}[h]
    \centering
    \includegraphics[scale = 0.5]{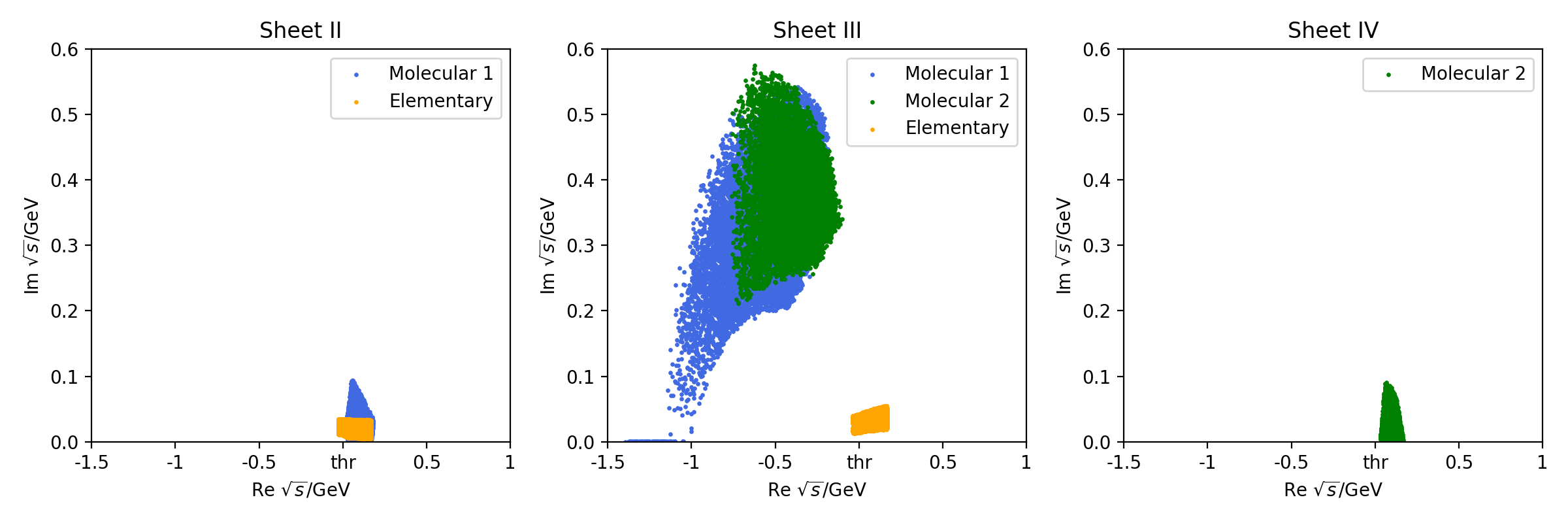}\\
    \caption{The pole locations of `molecule 1' (blue dots), `molecule 2' (green dots), and `elementary state' (yellow dots) of the training data. {Here `molecule 1' corresponds to a bound state of channel 2, whereas `molecule 2' corresponds to a virtual pole of channel 2.}}
    \label{sample}
\end{figure}

Considering that the experimental data is affected by the energy resolution, to match the real signal better, we add a gaussian convolution to the model. According to the experiment by LHCb~\cite{6900}, we fix $\Delta = 5~\mathrm{MeV}$ and
\begin{equation}
    \frac{d \sigma_i}{d \sqrt{s}} \sim \frac{1}{\sqrt{2 \pi} \Delta} \int_{\sqrt{s}-3 \Delta}^{\sqrt{s}+3 \Delta} \mathcal{R}_i\left(s^{\prime}\right) e^{-\frac{\left(\sqrt{s}-\sqrt{s^{\prime}}\right)^2}{2 \Delta^2}} d \sqrt{s^{\prime}}\ .
\end{equation}
According to above equation, when the parameter space is determined, all the training data can be  generated. Some pre-treatment is made to fit our machine learning program. The first is to set the window size of the signal.
 For FS1 di-$J/\psi$ signal, we set the window size from $E_{th2} - 100$ MeV to $E_{th2} + 300$ MeV to cover the relevant data and keep away from the noise from the peak X(6200), where $E_{th2}$ is the threshold energy for FS2. For $J/\psi \psi(2S)$  FS2 signal, we set the window size from $E_{th2}$ to $E_{th2} + 200$ MeV. See Figs.~\ref{ff1} and \ref{ff2} for two typical examples of the generated training data. The energy interval is fixed at 1 MeV uniformly so it can meet the demand that all the inputs sent to a special neural network should have the same size. The values at every energy point are calculated using a linear extrapolation of the experimental data, and their errors are taken randomly as 5, 10, 15 percent of their values. Finally, the normalization is made before sending them as learning input.
In Fig.~\ref{sample} examples of pole locations of two `molecules' and an `elementary' state are drawn and  are labeled appropriately using PCR.


Before sending the experimental data into the neural network, it is noted that the energy interval is wider than $1$ MeV. We have to manage these data to make the input the same size as the neural network. To be specific, if the energy interval of experimental
data is larger than 1 MeV then the method of linear interpolation will be employed to supply extra points.

\begin{figure}[H]
    \centering
    \subfigure[]{
    \includegraphics[scale = 0.385]{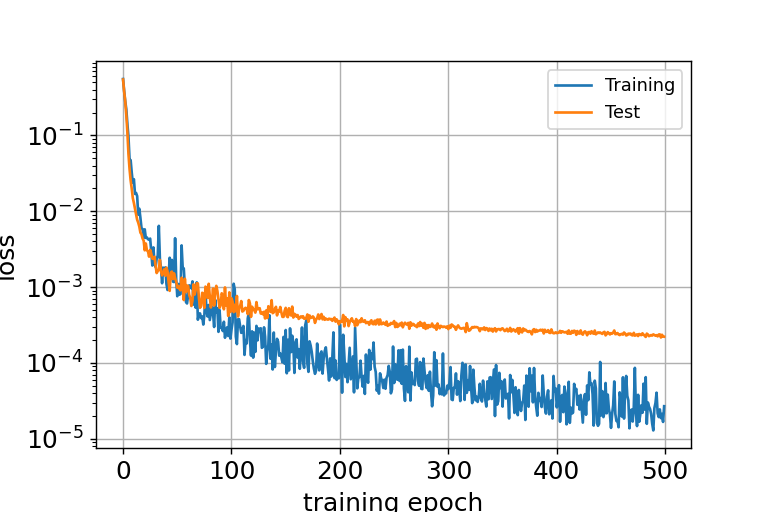}
    }
    \subfigure[]{
    \includegraphics[scale = 0.35]{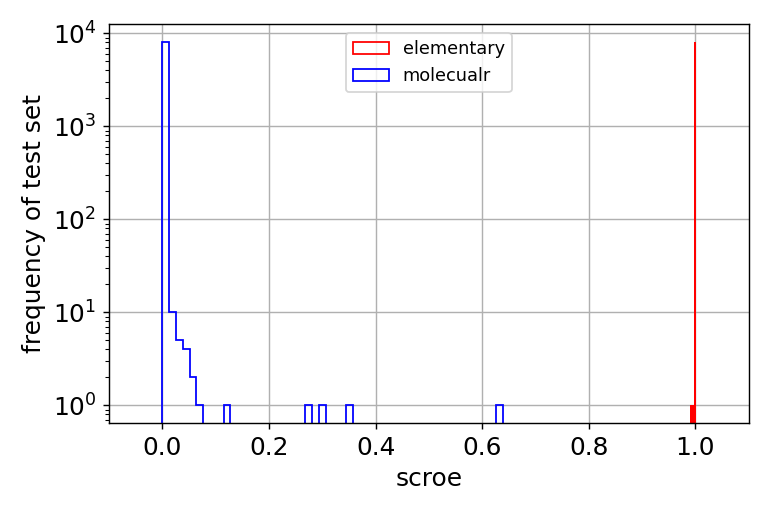}
    }
    \caption{left: loss function of the training set (blue), and test set (orange); right: the output result  of test set.    }
    \label{tulearn2}
\end{figure}

\begin{figure}[H]
    \centering
    \includegraphics[scale = 0.4]{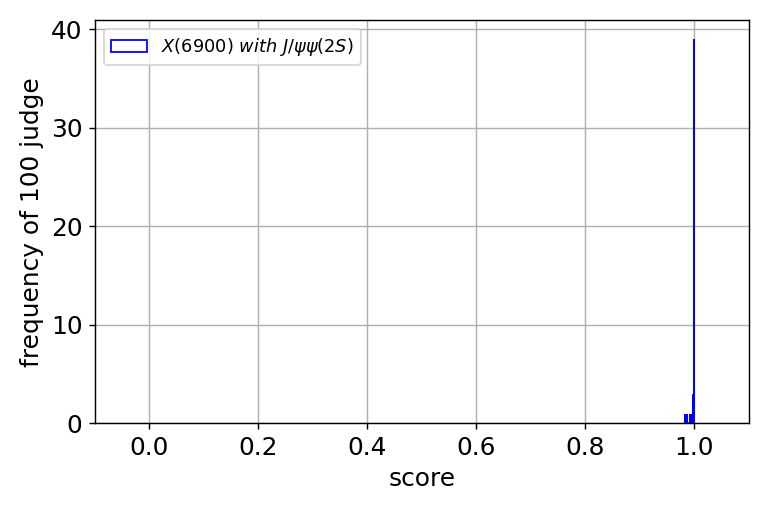}\\
    \caption{Test result of true data, it indicates that $X(6900)$ is not a molecule of $J/\psi\psi(2S)$.}
    \label{tutr}
\end{figure}

Fig.~\ref{tulearn2} shows the loss function of our two channel machine for both training data and test data. One can see that the data are trained well and can be used to judge whether a  peak stands for a molecule of a  given threshold.
As shown in Fig.~\ref{tutr}, the 100 data points got from the experimental data are all judged around `1' by the trained machine. It tells us that $X(6900)$ is not  a molecule composed of $J/\psi \psi(3686)$.

\textbf{A triple channel study:}

\begin{figure}[h]
    \centering
    \includegraphics[scale = 0.4]{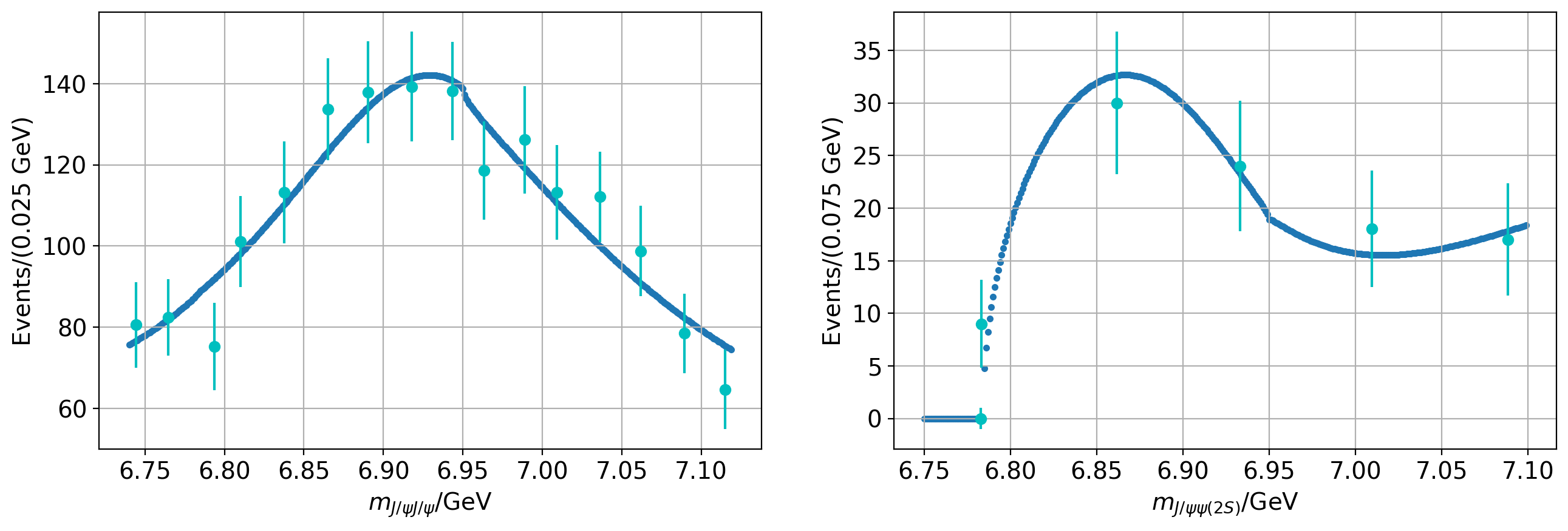}\\
    \caption{Fit to the  invariant mass spectrum, for the  `elementary solution' in table 1.
    The data are from Ref.~\cite{CMS}.}\label{fig3}
\end{figure}

\begin{figure}[h]
    \centering
    \includegraphics[scale = 0.4]{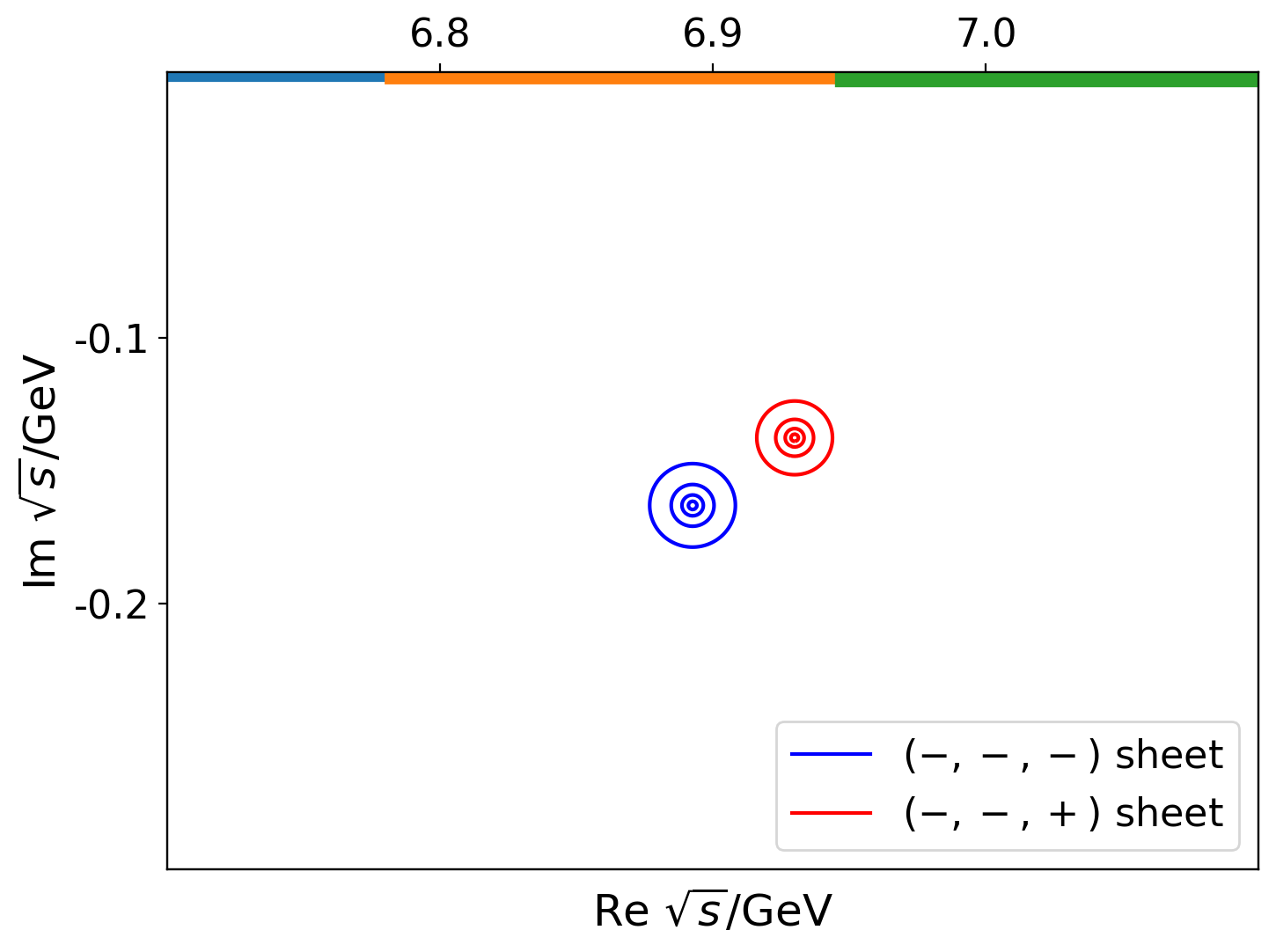}\\
    \caption{Pole positions of the `elementary solution' in table 1.}\label{fig3'}
\end{figure}

\begin{figure}[h]
    \centering
    \includegraphics[scale = 0.4]{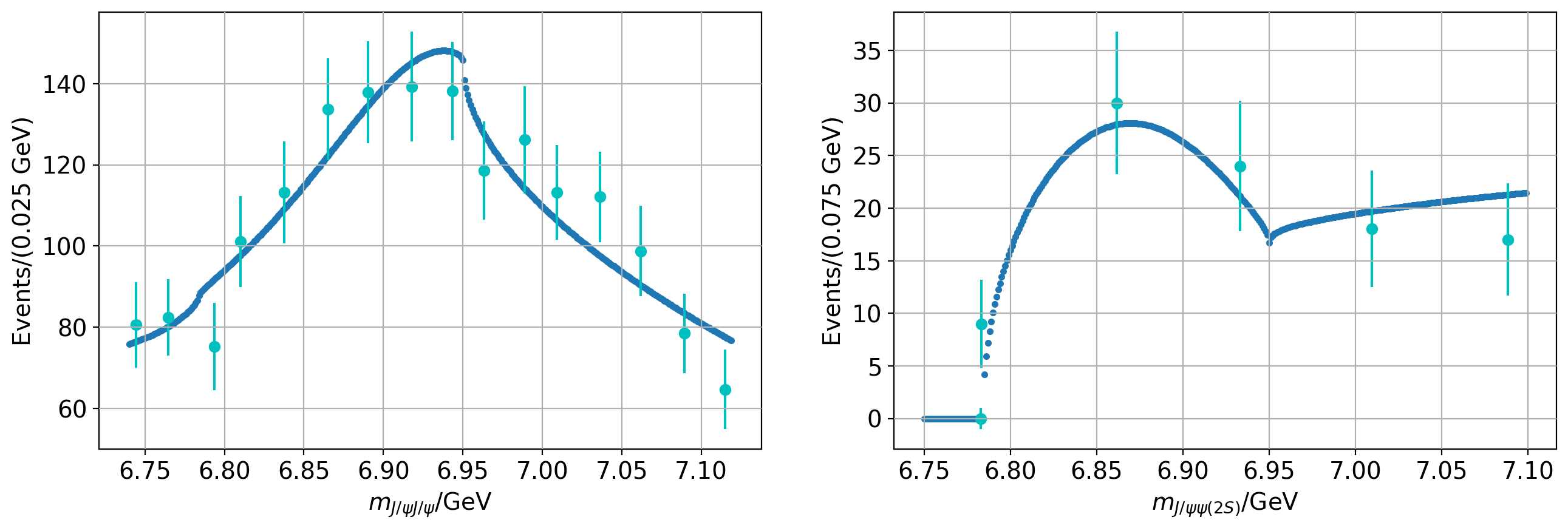}\\
    \caption{Fit to the  invariant mass spectrum, for the  molecule solution in table 1. The data are from Ref.~\cite{CMS}.}\label{fig4}
\end{figure}

\begin{figure}[h]
    \centering
    \includegraphics[scale = 0.35]{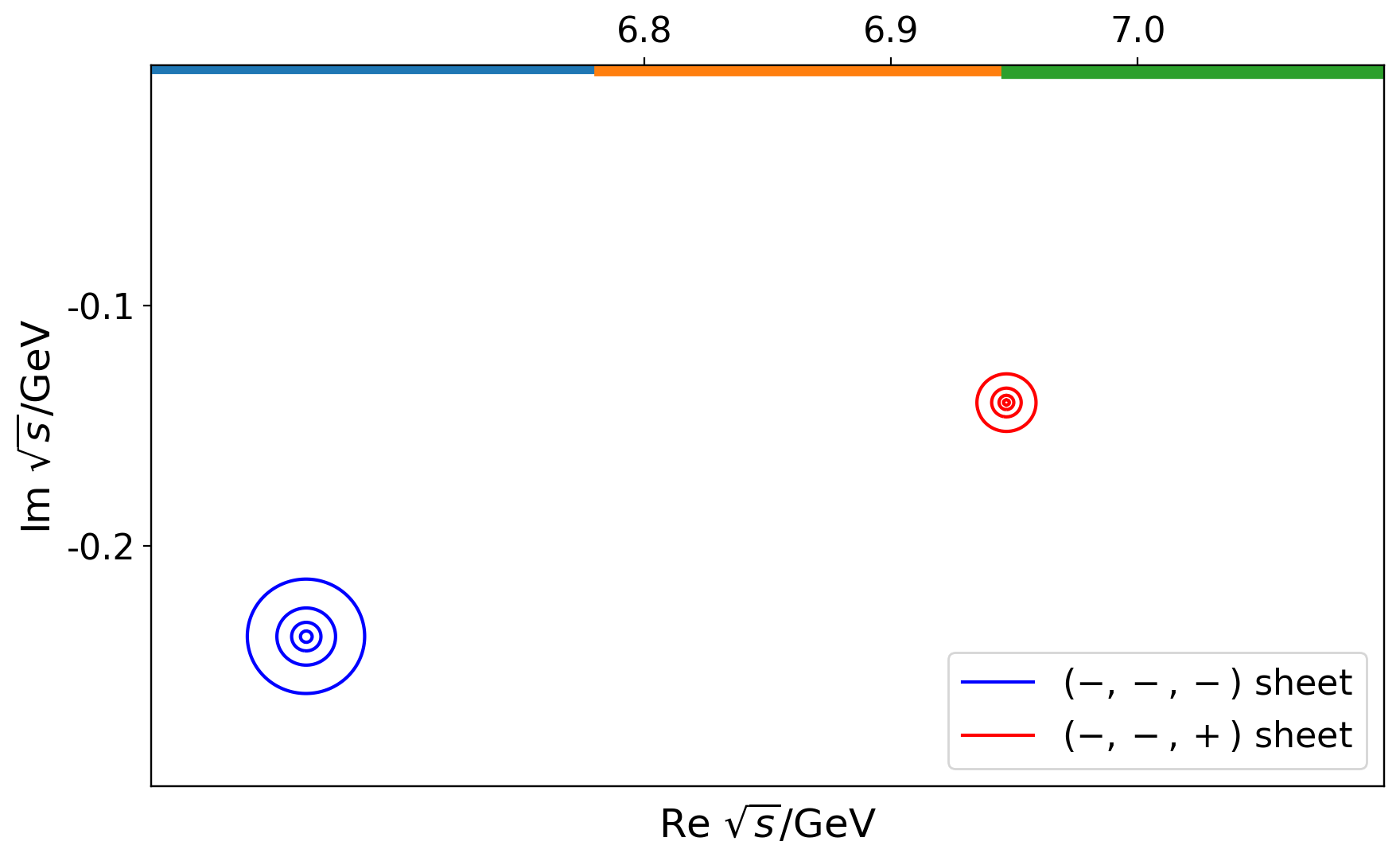}\\
    \caption{Pole positions of the molecule solution in table 1.}\label{fig4'}
\end{figure}

From above discussion,  it is concluded  that $X(6900)$ is not a molecule of $J/\psi \psi(2S)$, but it does not necessarily  mean that   $X(6900)$ is `elementary' since it may still be a molecule in some other higher channels, e.g.,  $J/\psi \psi(3770)$, $\chi_{c0}\chi_{c2}$, etc.
To examine this possibility, we extend the Flatt\'e-like parametrization to three channels:
\begin{equation}
    \begin{aligned}
    \mathrm{Event}_i(s) & =\mathrm{C}_i\rho_i(s)\left|\frac{e^{i \phi_i}}{s-M^2+i M\left(g_1 \rho_1(s)+g_2 \rho_2(s)+g_3 \rho_3(s)\right)}+{\left(a +b(\sqrt{s} - E_{thi})\right)}\right|^2.
    \end{aligned}
\end{equation}
In this part, we try to fit $J/\psi J/\psi$ and $J/\psi \psi(2S)$ signal directly by using above Flatt\'e-like parametrization equation. $E_{thi}$, $i = 1, 2$ are the threshold energies of FS1 and FS2.
All the parameters $g_1$, $g_2$, $g_3$, $\phi_i$, $a$, $b$, $\mathrm{C}_i$ are considered as fit parameters where $\mathrm{C}_i$ is a normalization constant. Here the `third channel' is taken as $\chi_{c0}\chi_{c2}$ for example (corresponding to $2^{++}$ assignment of $X(6900)$).\footnote{The $J/\psi\psi(3770)$  or $\chi_{c1}\chi_{c1}$ molecule ($0^{++}$ or $2^{++}$) are also possible, though slightly less preferred. In the former situation, $X(6900$ is a virtual state composed of $J/\psi\psi(3770)$.} As expected, the current data leave room for totally different scenarios. See  table~\ref{tab1}, apparently there exist multi-solutions. One is an `elementary' solution, which is similar to the couple channel solution discussed previously. However, when the coupling to the `third' channel, $g_3$, tuned large, a molecule (of the third channel) solution appears. See figs.~\ref{fig3}, \ref{fig3'} and figs~\ref{fig4}, \ref{fig4'} for illustration.
\begin{table}[H]
\centering
\begin{tabular}{lccccccccccc}
   \toprule
   Parameter& $g_1$ & $g_2$ & $g_3$ & M & a & b &$\phi_1$ & $\phi_2$ & $\mathrm{C}_1$ &$\mathrm{C}_2$& $\chi^2$/d.o.f\\
   \midrule
   Elementary &0.64  & 0.07 & 0.21 & 6.93 & 0.88& -0.39 &1.49&3.82&276&275&0.69 \\
   Molecule & 0.69 & 0.48 & 1 & 6.91 & 0.82 &-0.46&1.31&3.78&477&263&0.73 \\
   \bottomrule
\end{tabular}
\caption{Fit parameters for two different solutions.}\label{tab1}
\end{table}

Based on the standard PCR analysis we suggest  that it is possible the $X(6900)$ peak may be a molecule composed of, for example, $\chi_{c0}\chi_{c_2}$.
In principle, one may also make use of neural network to test the three channel situation. However in the absence of
data of the third channel, it is very difficult to make {a reliable test.}
 Finally we would like to comment on the work of Ref.~\cite{Zhou:2022xpd}, in which the authors also adopted a triple channel study (di-$J/\psi$, $J/\psi\psi(2S)$, $J/\psi\psi(3770)$). They conclude from their fit results (without including the new $J/\psi\psi(2S)$ data of Ref.~\cite{ATLAS:2022hhx}) that the $X(6900)$ is likely a true tetra quark state. However it is clear indicated from our analysis that the fit program contains redundancy of fit parameters and it certainly leaves the room for a `molecule' solution.
So additional experimental information in the  $\chi_{c0}\chi_{c_2}$, $J/\psi \psi(3770)$ channels, etc., are needed to further clarify the issue on whether $X(6900)$
is a molecular state or a compact tetra-quark state.

\textbf{Acknowledgements:}
This work is supported in part by National Nature Science Foundations
of China under Contract Number 11975028 and 10925522.


\begin{thebibliography}{}
%
% and use \bibitem to create references.
%
%------ Format for Journal Reference
%  \bibitem{RefJ}
%  Author, Journal \textbf{Volume}, (year) page numbers.
%------ Format for books
% \bibitem{RefB}
% Author, \textit{Book title} (Publisher, place year) page numbers



    \bibitem{6900}
       Roel Aaij \textit{et al.} (LHCb Collaboration), Sci.Bull. \textbf{65} (2020) 23, 1983.

    \bibitem{CMS} CMS collaboration, CMS-PAS-BPH-21-003 (2022).

    \bibitem{ATLAS:2022hhx} ATLAS collaboration, ATLAS-CONF-2022-040 (2022).




\bibitem{QFCao}
Q. F. Cao $et$ $al.$,
%Some remarks on X(6900)
Chin. Phys. \textbf{C} 45 (2021) 10, 103102.


\bibitem{Gong:2020bmg} C. Gong, M. C. Du, Qiang Zhao, X. H. Zhong, and B. Zhou, Phys. Lett. B \textbf{824} (2022) 136794.

\bibitem{Dong:2020hxe} X. Dong, F. K. Guo and B. S. Zou, Phys. Rev. Lett. \textbf{126} (2021) 152001.

\bibitem{Liu:2020tqy} M. Z. Liu and L. S. Geng, Eur. Phys. J. C \textbf{81} (2021) 2, 179.

\bibitem{Goncalves:2021ytq} V. P. Gon\c{c}alves and B.  Moreira, Phys. Lett. B \textbf{816} (2021) 136249.

\bibitem{Ke:2021iyh} H. W. Ke, X. Han, X. H. Liu and Y. L. Shi, Eur. Phys. J. C \textbf{81} (2021) 5, 427.

\bibitem{Liang:2021fzr} Z.  R. Liang, X. Y. Wu and D. L. Yao, Phys. Rev. D \textbf{104} (2021) 3, 034034.

\bibitem{Mutuk:2021hmi} H. Mutuk, Eur. Phys. J. C \textbf{81} (2021) 4, 367.

\bibitem{Wang:2021kfv} G. J. Wang, L. Meng, M. Oka, and S. L. Zhu, Phys. Rev. D \textbf{104} (2021) 3, 036016.

\bibitem{Wang:2021taf} Q. N. Wang, Z. Y. Yang, W. Chen, and H. X. Chen, Phys. Rev. D \textbf{104} (2021) 1, 014020.

\bibitem{Nefediev:2021pww} A. V. Nefediev, Eur. Phys. J. C \textbf{81} (2021) 8, 692.

\bibitem{Lu:2021kut} Q. F. L\"u, D. Y. Chen, Y. B. Dong and E. Santopinto, Phys. Rev. D \textbf{104} (2021) 5, 054026.

\bibitem{Chen:2021tad}  H. Chen, H. R. Qi and H. Q. Zheng, Eur. Phys. J. C \textbf{81} (2021) 9, 103102.

\bibitem{Wang:2021mma} Q. N. Wang, Z. Y. Yang and W. Chen, Phys. Rev. D \textbf{104} (2021) 11, 114037.

\bibitem{Esposito:2021ptx} A. Esposito, C. Andrea Manzari, A. Pilloni and A. D. Polosa, Phys. Rev. D \textbf{104} (2021) 11, 114029.

\bibitem{Liu:2021rtn} F. X. Liu, M. S. Liu, X. H. Zhong and Q. Zhao, Phys. Rev. D \textbf{104} (2021) 11, 116029.

\bibitem{Zhuang:2021pci} Z. J. Zhuang, Y. Zhang, Y. Z. Ma and Q. Wang, Phys. Rev. D \textbf{105} (2022) 5, 054026.

\bibitem{Chen:2021ftn} S. Z. Chen {\it et al.}, arXiv: 2111.14360 [hep-ph].

\bibitem{Lebed:2022gfb} R. F. Lebed,  Moscow Univ.Phys. Bull. \textbf{77} (2022) 2, 458.

\bibitem{Wang:2022jvk} Z. G. Wang, Int. J. Mod. Phys. A \textbf{37} (2022) 31n32, 2250189.

\bibitem{Wu:2022qwd} R. H. Wu {\it et al.}, JHEP \textbf{11} (2022) 023.

\bibitem{Chen:2022shj} C. Chen, H. Chen, W.~Q.~Niu, H.~Q.~Zheng, Eur. Phys. J. C \textbf{83}  (2023)1, 52.

\bibitem{Liang:2022rew} Z. R. Liang and De-Liang Yao, Rev. Mex. Fis. Suppl. \textbf{3} (2022) 3, 0308042.

\bibitem{Gong:2022hgd} C. Gong, M. C. Du and Q. Zhao, Phys. Rev. D \textbf{106} (2022) 5, 054011.

\bibitem{Chen:2022sbf} H. X. Chen, Y. X. Yan and W. Chen, Phys. Rev. D \textbf{106} (2022) 9, 094019.

\bibitem{Wang:2022yes} G. J. Wang, Q. Meng and M. Oka, Phys. Rev. D \textbf{106} (2022) 9, 096005.

\bibitem{Mutuk:2022nkw} H. Mutuk, Phys. Lett. B \textbf{834} (2022) 137404.


%\bibitem{Dong:2022sef} W. C. Dong and Z. G. Wang, arXiv: 2211.11989 [hep-ph].



%%%%%%%%%% After Cao Qin Fang 6900, new citations (above)





   \bibitem{BelleZb} A.~Bondar \textit{et al.} (Belle Collaboration),
   %``Observation of two charged bottomonium-like resonances in Y(5S) decays,''
   Phys. Rev. Lett. \textbf{108} (2012) 122001.

   \bibitem{BESIII3900} M.~Ablikim \textit{et al.} (BESIII Collaboration),
   %``Observation of a Charged Charmoniumlike Structure in $e^+e^-$ \textrightarrow{} $¹^+¹^-$ J/\ensuremath{\psi} at $\sqrt{s}$ =4.26  GeV,''
   Phys. Rev. Lett. \textbf{110} (2013) 252001.

   \bibitem{Belle3900} Z.~Q.~Liu \textit{et al.} (Belle Collaboration),
   %``Study of $e^+e^- ? ¹^+ ¹^- J/?$ and Observation of a Charged Charmoniumlike State at Belle,''
   Phys. Rev. Lett. \textbf{110} (2013) 252002.

   \bibitem{CLOE3900} T.~Xiao, S.~Dobbs, A.~Tomaradze and K.~K.~Seth,
   %``Observation of the Charged Hadron $Z_c^{\pm}(3900)$ and Evidence for the Neutral $Z_c^0(3900)$ in $e^+e^-\to \pi\pi J/\psi$ at $\sqrt{s}=4170$ MeV,''
   Phys. Lett. B \textbf{727} (2013) 366.

   \bibitem{Pc2015} R.~Aaij \textit{et al.} (LHCb Collaboration),
   %``Observation of $J/\psi p$ Resonances Consistent with Pentaquark States in $\Lambda_b^0 \to J/\psi K^- p$ Decays,''
   Phys. Rev. Lett. \textbf{115} (2015) 072001.

   \bibitem{Pc2019} R.~Aaij \textit{et al.} (LHCb Collaboration),
   %``Observation of a narrow pentaquark state, $P_c(4312)^+$, and of two-peak structure of the $P_c(4450)^+$,''
   Phys. Rev. Lett. \textbf{122} (2019) 222001.

 %  \bibitem{Cao:2020gul} Q. F. Cao, H. R. Qi, Y. F. Wang and H. Q. Zheng, Phys. Rev. D \textbf{100} (2019) 5.



   %%% pole counting
   \bibitem{pole} D.~Morgan, Nucl. Phys. A \textbf{543} (1992) 632.

   \bibitem{Zhang:2009bv} O.~Zhang, C.~Meng and H.~Q.~Zheng,
   %``Ambiversion of X(3872),''
   Phys. Lett. B \textbf{680} (2009) 453.

   \bibitem{Dai:2012pb} L.~Y.~Dai, M.~Shi, G.~Y.~Tang and H.~Q.~Zheng,
   %``Nature of X(4260),''
   Phys. Rev. D \textbf{92} (2015) 1, 014020.

   \bibitem{Cao:2019wwt} Q.~F.~Cao, H.~R.~Qi, Y.~F.~Wang and H.~Q.~Zheng,
   %``Discussions on the line-shape of the $X$(4660) resonance,''
  Phys. Rev. D \textbf{100} (2019) 5, 054040.



   \bibitem{X3900} Q.~R.~Gong, Z.~H.~Guo, C.~Meng, G.~Y.~Tang, Y.~F.~Wang and H.~Q.~Zheng,
   %``$Z_c(3900)$ as a $D\bar{D}^*$ molecule from the pole counting rule,''
   Phys. Rev. D \textbf{94} (2016) 11, 114019.



   %%% Spectral density function sum rule
   \bibitem{Baru:2003qq} V.~Baru, J.~Haidenbauer, C.~Hanhart, Y.~Kalashnikova and A.~E.~Kudryavtsev,
    %``Evidence that the a(0)(980) and f(0)(980) are not elementary particles,''
    Phys. Lett. B \textbf{586} (2004) 53.

    \bibitem{Weinberg} S.~Weinberg,
    %``Elementary particle theory of composite particles,''
    Phys. Rev. \textbf{130} (1963), 776.

    \bibitem{Weinberg:1965zz} S.~Weinberg,
    %``Evidence That the Deuteron Is Not an Elementary Particle,''
     Phys. Rev. \textbf{137} (1965), B672.

    \bibitem{Kalashnikova:2009gt} Y.~S.~Kalashnikova and A.~V.~Nefediev,
     %``Nature of X(3872) from data,''
      Phys. Rev. D \textbf{80} (2009) 074004.



\bibitem{Zhou:2022xpd} Q. Zhou, D. Guo, S. Q. Kuang, Q. H. Yang and L. Y. Dai, Phys. Rev. D \textbf{106} (2022) 11, L111502.


\end{thebibliography}
\end{document}